\documentclass[superscriptaddress, pre, twocolumn, aps, floatfix, 10pt]{revtex4-2}
\usepackage{amsmath}
\usepackage{amsfonts}
\usepackage{amssymb}
\usepackage{dsfont}
\usepackage{graphicx,graphics}
\usepackage{bm,bbm}
\usepackage{color}
\usepackage[colorlinks=true,linkcolor=blue,urlcolor=blue,citecolor=blue]{hyperref}
\usepackage[font=small,label font=bf]{caption}
\DeclareGraphicsExtensions {.pdf,.png,.jpg} 
\usepackage{subfig}
\usepackage{tikz}   
\usetikzlibrary{shapes}
\usepackage{tikz-cd}   
\usepackage{pgfplots}

\newcounter{example}[section]

\DeclareRobustCommand\dotOran{\tikz \fill[orange]          (0,0) circle (0.1);}

\DeclareRobustCommand\dotRed {\tikz \fill[red]             (0,0) circle (0.1);}
\DeclareRobustCommand{\dotMagenta}{\tikz {\fill[color=magenta] (0,0) circle (0.03); 
\node[thick, scale=0.6, circle, draw, color=magenta] at (0,0) {};}}

\DeclareRobustCommand\DiamGreen{\tikz \fill[black!50!green,rotate=45] (0,0) 
rectangle (0.15,0.15);}

\DeclareRobustCommand\DiamRed{\tikz \fill[red] (0,0) rectangle (0.15,0.15);}

\DeclareRobustCommand\TriaDViol{%
\tikz \fill[violet,scale=0.1] (-1,1) -- (1,1) -- (0,-1) -- (-1,1);}

\DeclareRobustCommand{\SqMagen}{%
\tikz {\fill[color=magenta] (0,0) circle (0.03);
\node[thick, scale=0.7, regular polygon, regular polygon sides=4, draw, color=magenta] at (0,0) {};}}

\DeclareRobustCommand{\PentaCyan}{%
\tikz {\fill[color=cyan] (0,0) circle (0.01);
\node[thick, mark size=3pt,color=cyan] at (0,0){\pgfuseplotmark{pentagon}};}}

\newcommand{\tr}{\mbox{tr}}
\newcommand{\la}{\langle}
\newcommand{\ra}{\rangle}

\newcommand{\ket}[1]{|#1\rangle}
\newcommand{\bra}[1]{\langle #1|}
\newcommand{\mc}[1]{\mathcal{#1}} 
\newcommand{\bs}[1]{\boldsymbol{#1}}
\newcommand{\J}{\mc{J}}
\newcommand{\one}{\mathds{1}}

\begin{document}
\title{From classical to quantum stochastic processes }

\author{Gustavo Montes}
\email{bboyfone1@gmail.com}
\affiliation{Departamento de F\'\i sica, Universidad de Guadalajara,
   Guadalajara, Jal\'\i sco, M\' exico.}
\affiliation{Benem\' erita Universidad Aut\' onoma de Puebla, Apartado Postal 
   J-48, Instituto de F\'\i sica, 72570, M\' exico.}
\affiliation{Quantum Biology Laboratory, Howard University, Washington DC 20059, USA.}

\author{Soham Biswas}
\email{soham.biswas@academicos.udg.mx}
\affiliation{Departamento de F\'\i sica, Universidad de Guadalajara,
   Guadalajara, Jal\'\i sco, M\' exico.}

\author{Thomas Gorin}
\email{thomas.gorin@academicos.udg.mx}
\affiliation{Departamento de F\'\i sica, Universidad de Guadalajara,
   Guadalajara, Jal\'\i sco, M\' exico.}

\begin{abstract}
In this paper for the first time, we construct quantum analogs starting from 
classical stochastic processes, by replacing
random “which path” decisions with superpositions of all paths. This procedure
typically leads to non-unitary quantum evolution, where coherences are
continuously generated and destroyed. In spite of their transient nature, these
coherences can change the scaling behavior of classical observables. Using the
zero temperature Glauber dynamics in a linear Ising spin chain, we find quantum
analogs with different domain growth exponents. In some cases, this exponent is
even smaller than for the original classical process, which means that
coherence can play an important role to speed up the relaxation process.
\end{abstract}

\maketitle

\section{Introduction}

Stochastic processes are extensively used to model different classical systems 
in many disciplines starting from physics~\cite{KaTay75,privman} to 
finance~\cite{finance}, chemistry~\cite{chemKamp} to biology~\cite{bio}, 
computer science~\cite{comsc}, etc. There are different types of stochastic 
processes which include random walks~\cite{rw}, L\' evy walks~\cite{levy}, 
Brownian motion~\cite{brn}, random fields~\cite{rndfld}, branching 
processes~\cite{brnch}, etc.

Since the ground breaking work in Refs.~\cite{Sud61,GoKoSu76,Lin76}, which 
provide a complete formal characterization of Markovian quantum processes, 
there has been a lot of work dedicated to map out the frontier between quantum 
and classical stochastic processes. On the quantum side, we find the 
superposition principle and coherence, whereas on the classical side there are
random choice and stochasticity. Starting from a unitary quantum 
process of a perfectly isolated system, measurements~\cite{qst} and/or the 
coupling to an environment~\cite{BrePet02} gradually replace coherent 
superpositions by stochastic mixtures. So the origin of 
stochasticity in the quantum process is the measurement theory itself. As 
argued in~\cite{Giulini96} this is how the classical world appears in quantum 
theory. Recent efforts to quantify ``coherence'' from the point of view of a 
resource theory~\cite{BaCrPl14,WinYan16,ChiGouPRL16} lead to the question, if 
there are ``non-classical'' tasks which can be performed only at the expense of 
this resource~\cite{q2clst,Che19}. In all these efforts, the starting point 
(focus) has always been on the initially unitary quantum process, eventually 
trying to find related classical stochastic processes.

By contrast, in this paper for the first time ever, we start from a given 
classical process and ask what are the quantum processes which can be converted 
to that classical process, by repeated projective measurements. We call these 
processes ``quantum extensions'', and we provide general guiding principles for 
their construction. Instead of starting from the unitary dynamics of the closed 
quantum system and considering the effect of the coupling to environmental 
degrees of freedom, we start from the classical stochastic process, where we 
replace random choices by superpositions. There are several measures of 
coherence~\cite{BaCrPl14,WinYan16,ChiGouPRL16} which quantify the amount of 
superpositions. Our approach allows to investigate 
systematically the different non-unitary quantum processes in the vicinity of a 
given classical process. It thereby sheds new light on the question to what 
extent quantum effects may be important in a variety of phenomena usually 
studied classically. Particularly important examples of high current interest 
can be found in biology~\cite{Marais18,Cao20,Kim21}.

In order to demonstrate that these quantum extensions may have 
qualitatively different and potentially useful properties, we consider the zero 
temperature quenching dynamics~\cite{glauber} of the one-dimensional Ising 
model. There, we analyze in detail two different quantum extensions. In one 
case, the relaxation process becomes generally faster, in the other case it 
becomes slower, than the original process. We show that small quantum effects 
can make certain physical processes more efficient, like the case of the 
relaxation process towards equilibrium. Such relaxation processes play a very 
important role for different cases in quantum biology as recently reported 
in~\cite{Cao20}.

\section{\label{DSM} Discrete stochastic Markov process}

\subsection{\label{CDSM} Classical discrete Markov process}
For simplicity, we restrict ourselves to stochastic Markov processes in
discrete time with a discrete state space $\{ \, |j\ra\, \}$. We can then 
describe the dynamics of the system by a sequence of stochastic transition 
matrices $\J(n)$, where $n\in\mathbb{N}_0$ denotes discrete time. These 
transition matrices fulfill a composition law according to which
$[\, \J(n)\; \J(n-1)\; \ldots\; \J(m)\, ]_{ij}$ is the probability to find the 
system in state $|i\ra$ at time $n$, provided the system was in state
$|j\ra$ at time $m-1$. The evolution of an ensemble of initial states, can be 
described by an evolving probability vector $\bm{p}(n)$, such that 
$\bm{p}(n) = \J(n)\, \bm{p}(n-1)$.

Similar to the classical case, a quantum stochastic Markov process 
$\Lambda_\mc{Q}$ can be defined as a sequence of quantum maps, 
$\Lambda_\mc{Q}(n)$. These must be completely positive, trace preserving, 
linear maps (CPTP-maps) in the space of density matrices~\cite{BZ,HZ}. 
Therefore, instead of an evolving probability vector, we now have a density 
matrix evolving in time. Similar to the classical case, we have the following 
composition law:
\begin{equation}
\varrho(n)= \Lambda_\mc{Q}(n) \circ \Lambda_\mc{Q}(n-1) \circ \ldots \circ 
   \Lambda_\mc{Q}(m)\; [ \varrho(m-1)\, ] \; .
\end{equation}
For the quantum description of a given classical Markov process we define the 
Hilbert space as the space of linear combinations of the classical states, 
$\{ \, |j\ra\, \}$, with the scalar product chosen in such a way that these 
states are orthonormal. Then, a diagonal density matrix $\bar\varrho(n)$, 
represents an ensemble of classical states, in exactly the same way as the 
above mentioned probability vectors. It is then easy to verify that the set of
operators
\begin{equation}
\left\{ K_{ij}(n) =  \sqrt{\J_{ij}(n)}\; |i\ra\la j|\, \right\}_{i,j}
\end{equation}
is a valid set of Kraus operators, 
defining a sequence of quantum channels 
$\Lambda_{\mathcal{J}}(n)$~\cite{BrePet02} such that
\begin{equation}
\bar\varrho(n)= \Lambda_{\mathcal{J}}(n)[\bar\varrho(n-1)]
 = \sum_{ij} K_{ij}(n) \bar\varrho(n-1) K_{ij}(n)^\dagger\; .
    \label{eq:apply_LJ}
\end{equation}
This completes the description of the classical stochastic process in the
quantum channel formalism. 

According to the resource theory of quantum 
coherence~\cite{BaCrPl14, WinYan16, Tank}, the classical maps $\Lambda_\mc{J}(n)$ 
are completely incoherent channels as their image is always a diagonal density 
matrix. In addition they are unable to detect any coherences, as their image
is independent of any non-diagonal elements. 

\subsection{Criteria for Quantum extension}

We are now interested in quantum processes $\Lambda_\mc{Q}(n)$, which may be 
considered as ``quantum extensions'' of the classical process 
$\Lambda_\mc{J}$. In practical terms, we require the quantum process
to reduce to the classical process, if it is observed (i.e. measured) 
sufficiently often. This leads us to the following formal definition:
\paragraph{Definition} A quantum process described by a sequence of 
quantum maps $\Lambda_\mc{Q}(n)$ is a ``quantum extension'' of the classical 
process described by $\Lambda_\mc{J}(n)$ if and only if
\begin{equation}
\forall n\quad :\quad \Lambda_\mc{J}(n) = \mc{P} \circ \Lambda_\mc{Q}(n) 
   \circ \mc{P} \; ,
\label{ValidQECondition}\end{equation}
where $\mc{P}$ denotes a complete measurement of the set of classical states
(i.e. the basis states $\{ |j\ra\}$).

\begin{figure}
\includegraphics[width=0.44\textwidth]{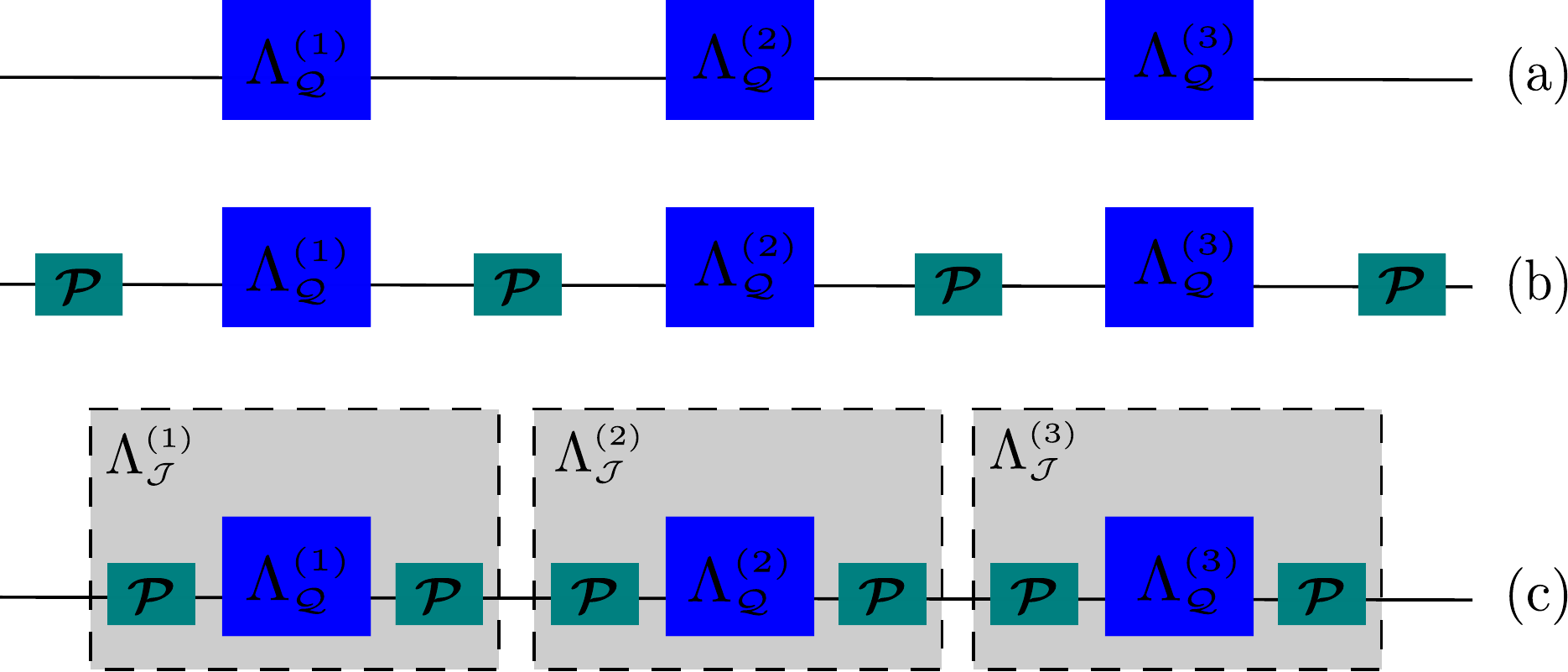}
\caption{Conversion of the three-time step quantum process $\Lambda_\mc{Q}$ 
consisting of the CPTP maps $\Lambda_\mc{Q}(n)$ shown in line (a) into
a classical process by complete measurements, denoted by $\mc{P}$. The result is
depicted in line (b). In line (c) 
the intermediate measurements are duplicated, which results in the classical 
maps $\Lambda_\mc{J}(n)= \mc{P}\circ\Lambda_\mc{Q}(n)\circ\mc{P}$. }
\label{f:QCdiag}\end{figure}

As illustrated in Fig.~\ref{f:QCdiag}, starting from a general 
quantum process $\Lambda_\mc{Q}$,
we include the complete measurements $\mc{P}$ before or after each individual 
quantum map $\Lambda_\mc{Q}(n)$.
The measurements between two quantum maps can be duplicated without changing 
the dynamics, which results in the classical maps 
$\Lambda_\mc{J}(n)= \mc{P}\circ\Lambda_\mc{Q}(n)\circ\mc{P}$.

To find candidate quantum processes $\Lambda_\mc{Q}$ which may then be checked
if they really are valid quantum extensions of a given classical process 
$\Lambda_\mc{J}$, we use the following guiding principles:
(i) We try to find quantum extensions for each stochastic map 
$\Lambda_\mc{J}(n)$, individually. 
(ii) Both, the classical and the quantum map can be uniquely defined by their
action on the basis states $\{ |j\ra \}$. Here, the classical map is determined 
by the probabilities to jump from the state $|j\ra$ to any other state. 
(iii) For the corresponding quantum map, we try to replace as far as 
possible random ``which path'' decisions by a superposition of all available 
options. This is done in such a way that the absolute value square of the 
transition amplitudes agree with the corresponding classical transition 
probabilities. 
 (iv) This must be done in such a way that 
Eq.~(\ref{ValidQECondition}) is fulfilled. As explained in App.~\ref{qsm} this 
is not always possible with a single unitary operation. Therefore we also 
consider conditional unitary operations, where the choice of the operation is 
conditioned on the outcome of a projective measurement. 
(v) Once we found a valid quantum extension 
$\Lambda_\mc{Q}(n)$, we may define a whole one parameter family of 
interpolating extensions by the convex sum 
\begin{equation}
     (1-p)\, \Lambda_\mc{Q}(n) + p\, \Lambda_\mc{J}(n)
    \label{Eq:interpolateQExt}
\end{equation}
for $0\le p \le 1$. Inserting Eq.~(\ref{ValidQECondition}) into
Eq.~(\ref{Eq:interpolateQExt}) shows that we can implement this map by simply
performing complete measurements $\mathcal{P}$ with probability $p$ after each 
elementary time step. Then for $p=0$ the system will be maximally coherent 
while for $p=1$ the process will be completely incoherent classical process.

\section{\label{IM} 
  Choice of the stochastic process and it's quantum extension}

To test our procedure for constructing quantum extensions of a given classical 
stochastic process, we consider the zero temperature quenching dynamics of the
one-dimensional nearest neighbor Ising model (without magnetic 
field)~\cite{Ising} with periodic boundary conditions. 

\subsection{\label{CIM} Classical relaxation process}

The classical relaxation process starts from a totally disordered initial 
configuration, corresponding to a very high temperature, where the spins 
are either up or down at random. We are interested in the relaxation process 
from this random initial configuration to one of the two ground states, where 
all spins are aligned to each other, pointing either up or down.

The relaxation process under zero temperature Glauber dynamics~\cite{glauber} 
is as follows: (i) choose a spin at random; (ii) calculate the energy change 
($\Delta E$) of the system if the spin is flipped; (iii) If $\Delta E < 0$ flip 
the spin, if $\Delta E > 0$ do nothing, if $\Delta E = 0$, choose one of the 
two options at random with equal probabilities. The process consisting of the 
steps (i),(ii),(iii) amounts to one elementary (local) operation. For a spin 
chain of length $N$, one Monte-Carlo step consists of $N$ such elementary 
operations.

 This process is a Markov chain~\cite{chemKamp}, exactly of the 
type considered in Sec.~\ref{CDSM}. If we denote the local operation, which is 
applied to spin $q$ in the chain by $\mc{J}^{(q)}$, then we find for a complete 
elementary operation:
\begin{equation}
\bm p(n+1) = \mc{J}\, \bm p(n) \; , \qquad 
\mc{J} = \frac{1}{N}\sum_{q=1}^N \mc{J}^{(q)} \; .
\label{M:defJall}\end{equation} 
Here, the uniform average taken over all local transition matrices 
$\mc{J}^{(q)}$, implements the random choice of spin $q$ in step (i), of 
the above list.

Finally, we will write down the transition matrix $\mc{J}^{(q)}$ in the basis
of all possible classical configurations. For that it is enough to consider
the spin $q$ and its immediate neighbors. Let us denote the spin-up state by 
$|0\ra$ and the spin-down state by $|1\ra$, and let us order the configurations
of three consecutive spins as $(\, |000\ra \, ,\; |001\ra \, ,\; |010\ra \, ,\; 
|011\ra \, ,\; |100\ra \, ,\; |101\ra \, ,\; |110\ra \, ,\; |111\ra \, )$. Then
the transition matrix reads
\begin{equation} 
\J^{(q)} = \begin{pmatrix} 
      1 &  0  & 1 &  0  &  0  & 0 &  0  & 0 \\
      0 & 1/2 & 0 & 1/2 &  0  & 0 &  0  & 0 \\
      0 &  0  & 0 &  0  &  0  & 0 &  0  & 0 \\
      0 & 1/2 & 0 & 1/2 &  0  & 0 &  0  & 0 \\
      0 &  0  & 0 &  0  & 1/2 & 0 & 1/2 & 0 \\
      0 &  0  & 0 &  0  &  0  & 0 &  0  & 0 \\
      0 &  0  & 0 &  0  & 1/2 & 0 & 1/2 & 0 \\
      0 &  0  & 0 &  0  &  0  & 1 &  0  & 1 
\end{pmatrix} \; . 
\label{M:defJ2}
\end{equation} 
In agreement with the Glauber process, outlined above, the transition matrix 
leaves the configurations $\ket{000}$ and $\ket{111}$ untouched, it flips the 
middle spin of the configurations $\ket{010}$ and $\ket{101}$, while flipping 
the middle spin with probability one half in all other cases. 

\subsection{Corresponding quantum extensions}

 In order to find quantum extensions of the process described in 
Sec.~\ref{CIM}, it is enough to find a quantum extensions $\Lambda_\mc{Q}$ of 
the classical transition matrix $\mc{J}$. Just as in the classical case, we 
will implement $\Lambda_\mc{Q}$ as a uniform mixture of local operations 
$\Lambda_\mc{Q}^{(q)}$, which are applied to spin $q$ and its immediate 
neighbors. Therefore,
\begin{equation}
\Lambda_\mc{Q} = \frac{1}{N}\sum_{q=1}^N \Lambda_\mc{Q}^{(q)}\; .
\end{equation} 
In this way, the problem is reduced to finding quantum extensions
$\Lambda_\mc{Q}^{(q)}$ for the three-spin classical operation $\mc{J}^{(q)}$. 

 As explained in detail in App.~\ref{qsm}, a single unitary 
operation cannot be a valid quantum extension of the map $\mc{J}^{(q)}$ given 
in Eq.~(\ref{M:defJ2}). One needs at least two Kraus operators, which may be 
chosen as conditional unitary operations. Dividing the three-spin Hilbert space 
into two subspaces, defined by the projectors, 
$\hat P = |010\ra\la 010| + |101\ra\la 101|$ and its complement 
$\hat P_{\rm c}$, we may write for the desired quantum operation
\begin{equation}
\Lambda_\mc{Q}^{(q)}[\varrho] = 
    \hat{\sigma}_x\, \hat P\, \varrho\, \hat P\, \hat{\sigma}_x
  + \hat U_X\, \hat P_{\rm c}\, \varrho\, \hat P_{\rm c}\, \hat U_X^\dagger\; .
\label{opsum}\end{equation}
Here, the Pauli matrix $\hat{\sigma}_x$ is applied to central spin, whereas the 
unitary matrix $\hat U_X$ can be decomposed into the orthogonal sum
\begin{equation}
\label{eq:3S_unitary}
\hat U_X = \one_{\{|000\ra,|111\ra\}} \oplus X_{\{ |001\ra,|011\ra\}} \oplus
  X_{\{ |100\ra,|110\ra\}}\; ,
\end{equation}
where $X$ may be any $2$$\times$$2$ dimensional unitary matrix with matrix 
elements of the same magnitude, $1/\sqrt{2}$. Again, in App.~\ref{qsm} we
explain how one arrives at this result and we show that $\Lambda_\mc{Q}^{(q)}$
is indeed a valid quantum extension, which fulfills 
Eq.~(\ref{ValidQECondition}).

In the numerical simulations to be presented below, we will focus on two
different choices for $X$, namely 
\begin{equation}
H = \frac{1}{\sqrt{2}}\begin{pmatrix} 1 & 1\\ 1 & -1\end{pmatrix}\; , \qquad
S = \frac{1}{\sqrt{2}}\begin{pmatrix} 1 & -1\\ 1 & 1\end{pmatrix}\; .
\label{M:HADSYH}\end{equation}
Here, $H$ is the Hadamard gate and $S$ is a SU$(2)$ matrix. As we will see,
the two options lead to the resulting quantum processes with very different 
behaviors.

\subsection{Observables}

In this section, we introduce the observables we are going to analyze. For 
studying the relaxation process in time, we measure time in units of 
Monte-Carlo time steps (MCS), such that $t$ increases by one unit, whenever $N$ 
elementary operations $\Lambda_\mc{Q}$ are completed.

\paragraph*{Average number of domain walls.}
For a classical Ising spin chain, starting from a disordered state, the average 
number of domain walls $D_w(t)$ decays as $t^{-1/z}$ indicating that the 
average domain size increases as $t^{1/z}$, where $z$ is domain growth 
exponent~\cite{bray}.

If $n_D(j)$ is the number of domain walls of the spin configuration $|j\ra$
($0\leq j \leq 2^N-1$), we define the corresponding quantum 
observable $\hat D_{\rm W} = \sum_j n_{\rm D}(j)\, |j\ra\la j|$. Hence
the expectation value for an evolving mixed quantum state $\varrho(t)$ is
$\la\hat D_{\rm W}\ra(t) = {\rm tr}\big [ \hat D_{\rm W}\, \varrho(t)\big ]$.

\paragraph*{Relaxation time}
The dynamics of the classical process always ends in a state,
where all spins point in the same direction, either up,
$\ket{\bs{0}}= \ket{00\cdots 0}$, or down, $\ket{\bs{1}}=\ket{11\cdots 1}$.
The quantum process, however, can also end in a superposition of the two 
states. Thus, the probability to find the system after time $t$ at equilibrium, 
is $P_{\rm Eq}(t) = {\rm tr}\big [ \hat P_{\rm Eq}\, \varrho(t)\big ]$, where 
$\hat P_{\rm Eq} = \ket{\bs{0}}\bra{\bs{0}} + \ket{\bs{1}}\bra{\bs{1}}$.
In the simulations below, we report the half life $t_{1/2}$ in MCS,
which is such that $P_{\rm Eq}(t_{1/2}) = 1/2$ (see App.~\ref{rlxt} for further elaboration). 

\paragraph*{Coherence}
Quantum extensions differ crucially from their classical ancestor due to the
presence of superpositions. These are quantified conveniently with the help
of the coherence measure
\begin{equation}
C(t) = \sum_{i\ne j} |\varrho_{ij}(t)| \; ,
\label{l1Coherence}\end{equation}
as introduced in~\cite{BaCrPl14}.

\section{Numerical simulations}

We simulated different quantum extensions of the classical relaxation process 
for the zero temperature quench of an Ising spin chain. At their core there
are the unitary matrices $H$ and $S$ [Eq.~(\ref{M:HADSYH})], which generate 
the required quantum superpositions. We label the resulting processes by 
``HAD'' and ``SYH'', respectively. 
As explained above, once a quantum extension $\Lambda_\mc{Q}$ is found, we can 
construct convex combinations of that process with the original classical 
process as described in Eq.(\ref{Eq:interpolateQExt}).This allows 
us to control the degree of coherence introduced in the system.

In practice this means that after each elemental quantum map 
$\Lambda_\mc{Q}^{(q)}$, we apply a complete measurement $\mc{P}$ with 
probability $p$. We consider three cases: $p=0$ 
(maximally coherent without complete measurements), $p = 1/N$ (on average one 
complete measurement per MCS), $p=1$ (complete measurements with certainty 
reproducing the classical process). In that latter case, the two options 
``HAD'' and ``SYH'' yield exactly the same dynamics. 
This leaves us with five different processes listed in 
Tab.~(\ref{tab:interpolating_cases}).
\begin{table}[t]
    \centering
    \caption{Interpolation of the quantum extensions by means of the parameter
    $p$, as described in Eq.~(\ref{Eq:interpolateQExt}). The first row list the
    different processes considered here. The second row shows the value of $p$,
    which depends of the average number of projective measurements. The last row
    specifies the number of these projective measurements per MC step for each 
    of the defined processes.}
    \label{tab:interpolating_cases}
    \begin{tabular}{c|ccccc}
        \hline\hline
                      &            &            &            &            &            \\
        case:         & HAD-0\hspace{0.3cm} & HAD-1\hspace{0.3cm} & Class\hspace{0.3cm} & SYH-1\hspace{0.3cm} & SYH-0\hspace{0.0cm} \\\hline
        $p$           &   0        & $1/N$      &   1        & $1/N$      &   0        \\
        measure. rate &   0        &   1        &  $N$       &   1        &   0        \\
        \hline\hline
    \end{tabular}
\end{table}

\paragraph*{Methodology} 
In all cases, we consider an even number of spins $N$, with random initial 
configurations, where there are exactly as many spins pointing upward as 
downward. To avoid the evolution of huge density matrices, we use an 
unraveling method~\cite{BrePet02, stsch}. 
That is, we compute the evolution of a large number of pure states,
$\{ \Psi_\alpha(n) \}_{1\le \alpha \le N_{\rm sam}}$, and approximate the 
density matrix at multiples of the MCS as
\begin{equation}
\varrho(t) \approx \frac{1}{N_{\rm sam}}\sum_\alpha 
   |\Psi_\alpha(Nt)\ra\la\Psi_\alpha(Nt)| \; ,
\label{N:unravelling}\end{equation}
where $t$ is time in MCS. In the present case, the unraveling
is easy to realize, since the whole process has been decomposed into random
choices, measurements and conditional unitary operations [see App.~\ref{unrav} 
for details]. Using a sparse matrix storage format for the density matrices, 
We have been able to perform simulations up to $N=20$ spins averaging over
$N_{\rm sam}= 24\, 000$ random initial configurations~\cite{siml}.

\begin{figure}
\includegraphics[width=0.48\textwidth]{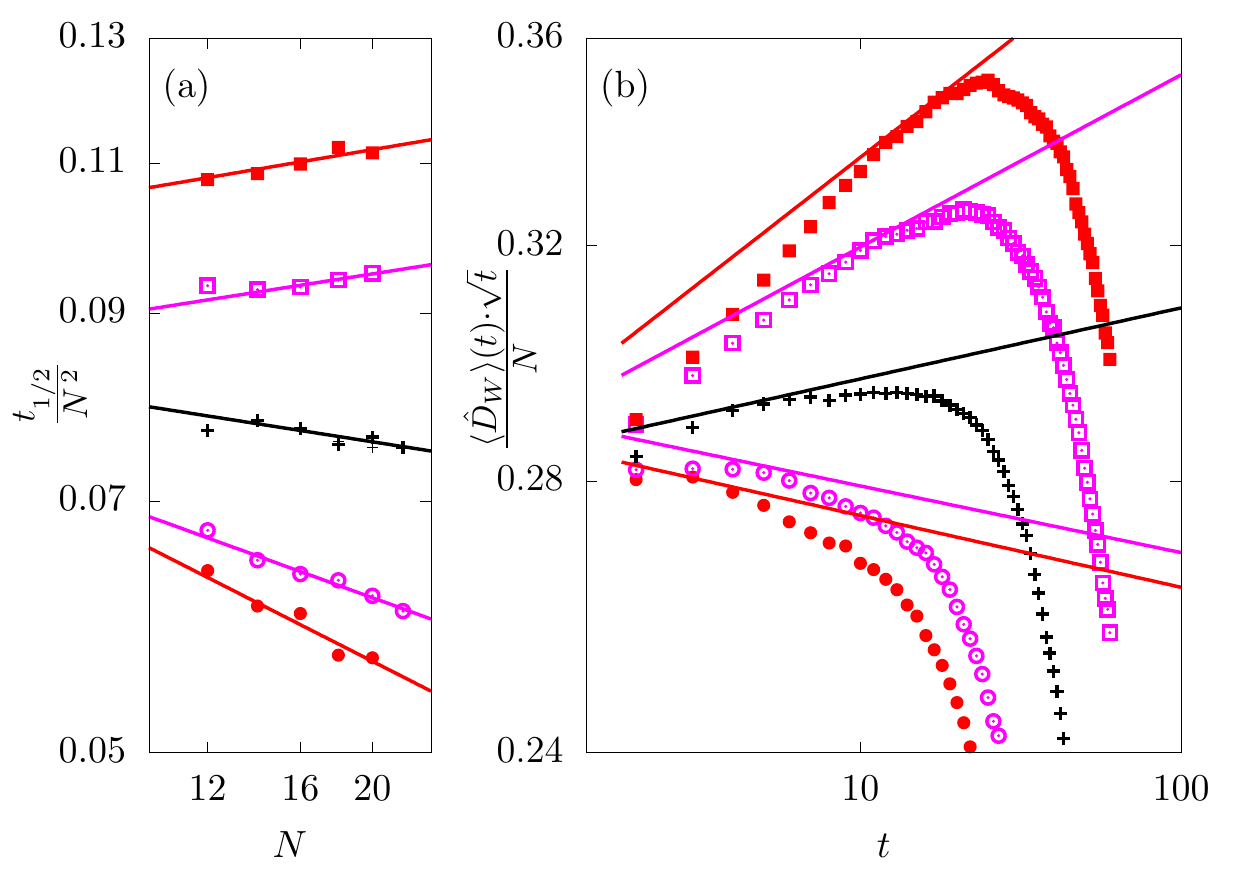}
\caption{Panel (a): half life $t_{1/2}$ (in units of MCS) divided by $N^2$ vs. 
$N$ (the length of the spin chain) in a log-log plot. Points are numerical 
simulations, the straight lines best fits. Panel (b): Average number of domain 
walls $\la\hat D_{\rm W}\ra$ multiplied by $\sqrt{t}/N$ vs. $t$ for the 
different process types for $N=20$. In both panels the symbols 
correspond to: HAD-0 ($\DiamRed$), HAD-1($\SqMagen$), Class($+$), SYH-1 ($\dotMagenta$), and SYH-0 ($\dotRed$).}
\label{Fig:Domains}
\end{figure}

On panel (a), Fig.~(\ref{Fig:Domains}) shows the half-time $t_{1/2}$ of the 
relaxation process as a function of the number of spins $N$. In the plot, we
divide $t_{1/2}$ by $N^2$ which corresponds to the scaling of the relaxation 
time for the classical process as $N\to\infty$~\cite{bray}. We show the results 
for the five different processes, described in 
Tab.~\ref{tab:interpolating_cases}. 
The finite negative slope in the classical case, 
seems to indicate that $t_{1/2} \sim N^z$ with $z < 2$. We checked that this is
due to the small system size. For $N \gtrsim 80$ one recovers the expected 
value $z \approx 2$. 
The quantum extensions show very different scaling exponents. For the HAD cases 
the exponents are larger, for the SYH cases smaller than in the classical case. 
These findings are remarkable, in view of the fact that the coherence which is 
built up during the process is transient and from a global perspective 
relatively week. The apparent acceleration of the relaxation process is also 
remarkable; in particular as it seems to persist in the large $N$ limit.

On panel (b), the figure shows the decay of the average number of domain walls 
with time for the largest spin chain ($N=20$). Here, $\la\hat D_{\rm W}\ra$ is
multiplied by $\sqrt{t}/N$ which would compensate the expected classical, 
large-$N$, power-law decay. In that way, data which are showing that power-law 
decay with $z=2$ would lie on a horizontal line. On general grounds the scaling 
coefficients for the half times in panel (a) are related to the scaling 
coefficients for the decay of $\la\hat D_{\rm W}\ra$ as $t_{1/2}\sim N^z$ and 
$\la\hat D_{\rm W}\ra \sim t^{-1/z}$~\cite{bray}. Thus, the solid lines in 
panel (b) show the power-laws obtained from the exponents estimated in panel 
(a).

For a realistic macroscopic relaxation process, the natural unit of time is 
the time it takes for a macroscopic number of constituents to change their
state. In our simulations, this time scale is the MCS. So one could expect that
if one removes all coherences at every MCS, that the macroscopic observables
should behave essentially as in the classical process. Our results for the 
cases HAD-1 and SYH-1 show that this is not the case. Both, the half times 
plotted in panel (a) and the average number of domain walls plotted in panel 
(b) are closer to the maximally quantum cases (without the complete projective
measurements) than to the classical case. This quite proves that provided the
generation of coherence is sufficiently effective even extremely short 
decoherence times will not be sufficient to suppress the particular properties
of the quantum process.

\begin{figure}
\includegraphics[width=0.48\textwidth]{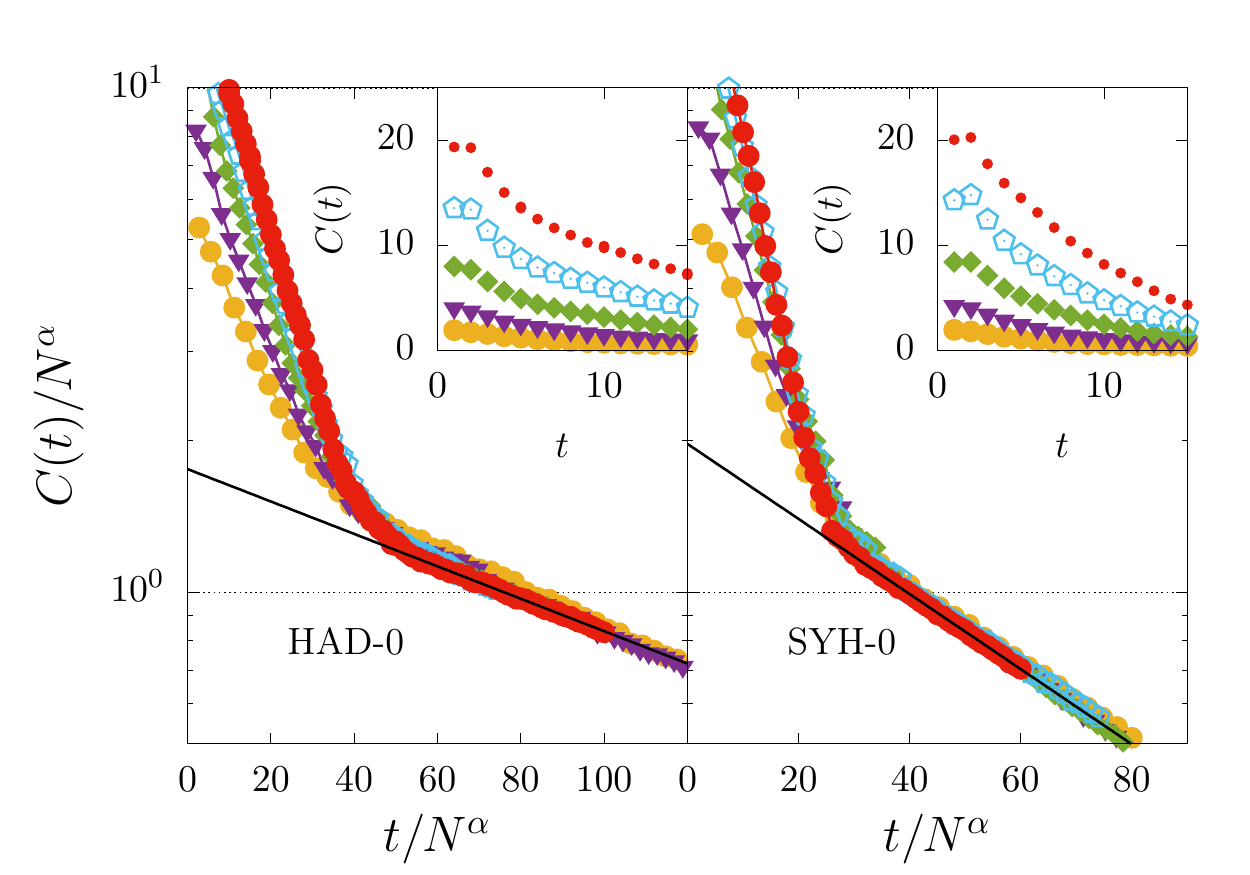}
\caption{Coherence measure from Eq.~(\ref{l1Coherence}) as a function of time
for different chain lengths, $N= 12\, (\dotOran)$; 
14 $(\TriaDViol)$; 16 $(\DiamGreen)$; 18 $(\PentaCyan)$; 20 $(\dotRed)$. 
The inset shows the original data, for short times. The main plot shows the 
data collapse at large times following the scaling law given in 
Eq.~(\ref{scl}). }
\label{f:Cohe}\end{figure}

Figure~\ref{f:Cohe} shows the coherence measure from Eq.~(\ref{l1Coherence}) as
a function of time, for different chain lengths.
Initially there is a steep increase with a pronounced maximum after a few MCS, 
followed by an apparently exponential decay. The maximum value of $C(t)$ 
increases with $N$, but slower than exponential. Since the number of 
non-diagonal elements of $\varrho(t)$ is $O(2^N)$, this indicates 
that the overall amount of coherence is rather small.
However as we have shown in figure \ref{Fig:Domains}, the amount of coherence
is enough to change the behavior of the other observables 
significantly. The scaled semi-log plot in the main figure, shows that the 
coherence decays exponentially for all times, with different decay rates at
short and large times. For large $t$, a scaling form can be written as
\begin{equation}
C(t) \sim N^{\alpha}\, \exp(k\, t/N^{\alpha}), 
\label{scl}
\end{equation}
where $\alpha = 1.91 \pm 0.014$, $k \simeq 0.017$ for SYH-0, and
$\alpha = 2.16 \pm 0.02$, $k\simeq 0.007$ for HAD-0.
The scaling form of $C(t)$ is more complex for short times \cite{unpub}. 

One could have assumed that coherence in general might speed up the 
relaxation process for similar reasons than in the case of the Grover
quantum search algorithm \cite{Grover}. Instead of applying the 
energy-minimizing spin-flip operations to individual configurations 
sequentially, coherence allows to apply these operations in parallel to several 
configurations. Our results show that the situation is more complicated. 
Almost the same amount of coherence (at the early stage of the process) 
can either lead to a speed-up or a slowing down of the relaxation process, 
depending on nothing else but the unitary spin operations employed (HAD or 
SYH). Interestingly, in the HAD case, the system not only takes a much longer 
time to reach the equilibrium subspace, it also conserves coherence for a much 
longer time. Hence, one can say that it is the conservation of coherence which 
makes it more difficult for the system to find its way to the equilibrium subspace. Of 
course it would be very desirable and in principle it should be possible to 
discover the mechanism behind such different behaviors. This requires more 
detailed studies of the relaxation process itself, which will be left as a task 
for the future.

\section{Conclusions and Discussion}
Through a new concept that we called quantum extension, we find quantum 
processes associated with a given Markovian classical stochastic process. In 
physical terms, this association means that sufficiently frequent observations 
of the quantum process will convert those processes into the original classical 
one. This construction does not need an auxiliary system~\cite{BZ,HZ}. In other
words, unlike in the case of quantum walks~\cite{qw}, our construction does not 
extend the state space of the original classical process~\cite{rw}.

We apply this new approach to the Glauber relaxation process of a linear Ising 
chain under a quench to zero temperature. We characterize the coherence of the 
system in different temporary regimes. Finally, we analyze the process to reach 
the equilibrium subspace, providing strong evidence that the relaxation times 
scale differently, for different quantum extensions. Among those, we find at 
least one process which is faster than the classical case.
 For a long time stochastic dynamics has been studied for 
several systems classically including the spin systems for which the real world 
is quantum. Our construction of ``quantum extensions'' is useful for studying 
the dynamics of such systems in a more realistic way.

Our findings may be particularly relevant in areas such as quantum 
biology and quantum computation. 
There are increasing experimental evidences~\cite{Marais18,Cao20,Kim21} for the 
presence of coherence in biological processes. Currently the main question is 
to what extent this coherence is necessary for the functioning and efficiency 
of these processes. In this context, our results (in particular those for
HAD-1 and SYH-1) show that even very short coherence times can be compensated 
by the continuous generation of superpositions, in such a way that the 
relaxation of macroscopic observables is strongly changed. The application of
our approach to such biological processes, may therefore shed new light on
these questions. 

In the field of quantum computations, quantum annealers~\cite{Wein20,Junger21} 
are designed to find the energetic minima of Ising-type Hamiltonians. However 
the general purpose of quantum computers are to implement idealistic unitary 
algorithms with the help of quantum error correction at the expense of an 
enormous number of additional qubits~\cite{Reiher17,Chao2018}. In this area, 
our approach could pave the way to form quantum operations which are partially 
incoherent, but still conserve/generate sufficient coherence to outperform a 
classical algorithm.

\section{acknowledgements}
The authors acknowledge the use of Leo-Atrox supercomputer at the CADS 
supercomputing center. This work has been financially supported by the Conacyt 
project ``Ciencias de Frontera 2019'', number 10872.

\appendix

  \renewcommand{\tr}{\mbox{tr}}
  \renewcommand{\la}{\langle}
  \newcommand{\kb}[2]{|#1\rangle\langle #2|}
   \newcommand{\prj}[1]{|#1\rangle\langle #1|}
  \newcommand{\braket}[2]{\langle #1 | #2 \rangle}
   \newcommand{\LQ}{\Lambda_{\mc{Q}}}
   \newcommand{\LJ}{\Lambda_{\mc{J}}}
  
\section{Quantum stochastic map} 
\label{qsm}

In order to find quantum extensions $\Lambda_\mc{Q}$ of a classical stochastic 
operation $\mc{J}$, we use the Kraus operator representation~\cite{kraus1983}. 
In that case, $\Lambda_\mc{Q}$ is defined in terms of a set of Kraus operators 
as
\begin{equation}
    \Lambda_\mc{Q}\quad :\quad \varrho \to \varrho' = \sum_r \hat{\mc{K}}_r\, \varrho\, 
    \hat{\mc{K}}_r^\dagger \; , \quad \sum_r \hat{\mc{K}}_r^\dagger \hat{\mc{K}}_r = \one \; .
\end{equation}
It is a very difficult task to systematically analyze all possible quantum 
extensions in terms of such general Kraus operators. We therefore consider
a more restrictive set of quantum operations, namely conditional unitary 
operations. In physical terms, that amounts to a projective measurement, with
subsequent unitary operations which are conditioned on the measurement 
outcomes.  In other words, for an observable $\hat A$ with the spectral 
decomposition
\begin{equation}
\hat A= \sum_r a_r\; \hat P_r \; , \quad 
\hat P_r\, \hat P_s = \delta_{rs}\, \hat P_r\; , \quad
\sum_r \hat P_r = \one\; ,
\end{equation}
and arbitrary unitary operators $U_r$, we choose the Kraus operators as
\begin{equation}
\hat{\mc{K}}_r = \hat{U}_r\,\hat{P}_r\, \; .
\label{eq:KrausOperators}\end{equation}

A case of particular interest is the case of a completely unitary operation,
when $\LQ$ contains only one element, 
$\hat{\mc{K}}_1=\hat{U}$ ($\hat{P}_1=\one$).
In this case the condition for a valid quantum extension, 
Eq.~(\ref{ValidQECondition}), requires that 
$\Lambda_\mc{J} = \mc{P} \circ \Lambda_\mc{Q} \circ \mc{P}$. For an arbitrary
density matrix $\varrho$ and $\varrho' = U\, \varrho\, U^\dagger$, this is
equivalent to the condition that
\begin{align}
    \label{eq:lhs_QEcondition}
    \LJ :\quad \varrho'_{jj}
    &= \sum_k \mc{J}_{jk}\varrho_{kk} \\ \,. 
    \label{eq:rhs_QEcondition}
    \mc{P}\circ\LQ\circ\mc{P} :\quad \varrho'_{jj}
    &= \sum_k\bra{j}\,\hat{U}\,\prj{k}\,\varrho\,\prj{k}\,\hat{U}^\dagger\,\ket{j} \nonumber \\
    &= \sum_k |U_{jk}|^2\, \varrho_{kk}\,,
\end{align}

and when equations (\ref{eq:lhs_QEcondition}) and (\ref{eq:rhs_QEcondition}) are
compared, the following condition is obtained
\begin{align}
    \label{eq:QE_unitaryCondition}
    \forall\, j,k : \qquad \mc{J}_{jk} = |U_{jk}|^2\,. 
\end{align}
For our particular example of the classical map defined in Eq.(\ref{M:defJ2}), 
such a unitary operation cannot exist. This is because the required unitary 
matrix must have matrix elements $U_{11}$ and $U_{13}$ of magnitude equal to 
one, so that the corresponding columns cannot be orthogonal.

If a single unitary operation does not work, we can try to find a quantum 
operations with two Kraus operators. As explained above, we search for Kraus 
operators which describe a conditional unitary operation. That means, depending
on whether the system is found in one subspace $\hat P$ or not (subspace 
$\hat P^{\rm c}$) we apply one unitary operation $\hat{U}$ or another, $\hat V$. To
simplify the task even more, we assume that the basis states are eigenstates
of $\hat P$ and $\hat P^{\rm c}$. With this, the requirement for a valid
operator sum representation is fulfilled:
\begin{align}
\hat{\mc{K}}_1 = \hat{U}\, \hat P\; , \quad \hat{\mc{K}}_2 = \hat{V}\, \hat P^{\rm c}   \nonumber \\
 \textrm{with}\quad 
\hat{\mc{K}}_1^\dagger \hat{\mc{K}}_1 + \hat{\mc{K}}_2^\dagger \hat{\mc{K}}_2 = \hat P + \hat P^{\rm c} = \one\,.
\end{align}
Now, from the right-hand side of equation (4), the following diagonal elements 
of $\varrho'$ are obtained
\begin{align}
    {\varrho}_{jj}' &= U_{jk}\, \hat P_{kk}\, \varrho_{kk}\, \hat P_{kk}\, U_{jk}^* + 
   V_{jk}\, \hat P^{\rm c}_{kk}\, \varrho_{kk}\, \hat P^{\rm c}_{kk}\, V_{jk}^* 
\end{align}
and the requirement for a valid quantum extension now reads:
\begin{align}
    \forall j,k :\quad J_{jk} = \begin{cases} 
   |U_{jk}|^2 &: \hat P\, |k\ra = |k\ra \\
   |V_{jk}|^2 &: \hat P\, |k\ra = 0\end{cases}\; . 
    \label{QE_2KrausElem}
\end{align}

\begin{widetext}
Now let us  return to our particular case. A quantum extension associated to 
    the classical map Eq.~(\ref{M:defJ2}) that satisfies condition~(\ref{QE_2KrausElem})
    is defined in Eq.~(\ref{opsum}) where the unitary operations $U$ and $V$ 
    are chosen as $\hat{U} = \hat{\sigma}_x^{(q)}$, $\hat{V} 
= \hat{U}_X$ (defined in Eq.~\ref{eq:3S_unitary}), which in matrix form look 
    like
\begin{align}
    \mc{K}_1 &= \sigma_x^{(q)}\,  P = 
\begin{pmatrix}
 0 & 0 & 1 & 0 & 0 & 0 & 0 & 0\\
 0 & 1 & 0 & 0 & 0 & 0 & 0 & 0\\
 1 & 0 & 0 & 0 & 0 & 0 & 0 & 0\\
 0 & 0 & 0 & 1 & 0 & 0 & 0 & 0\\
 0 & 0 & 0 & 0 & 1 & 0 & 0 & 0\\
 0 & 0 & 0 & 0 & 0 & 0 & 0 & 1\\
 0 & 0 & 0 & 0 & 0 & 0 & 1 & 0\\
 0 & 0 & 0 & 0 & 0 & 1 & 0 & 0\end{pmatrix}
\begin{pmatrix}
 0 & 0 & 0 & 0 & 0 & 0 & 0 & 0\\
 0 & 0 & 0 & 0 & 0 & 0 & 0 & 0\\
 0 & 0 & 1 & 0 & 0 & 0 & 0 & 0\\
 0 & 0 & 0 & 0 & 0 & 0 & 0 & 0\\
 0 & 0 & 0 & 0 & 0 & 0 & 0 & 0\\
 0 & 0 & 0 & 0 & 0 & 1 & 0 & 0\\
 0 & 0 & 0 & 0 & 0 & 0 & 0 & 0\\
 0 & 0 & 0 & 0 & 0 & 0 & 0 & 0\end{pmatrix} \; , 
\end{align}
\begin{align}
    \mc{K}_2 &= U_X\,  P^{\rm c} = 
\begin{pmatrix}
 1 & 0 & 0 & 0 & 0 & 0 & 0 & 0\\
 0 & X_{11} & 0 & X_{12} & 0 & 0 & 0 & 0\\
 0 & 0 & 1 & 0 & 0 & 0 & 0 & 0\\
 0 & X_{21} & 0 & X_{22} & 0 & 0 & 0 & 0\\
 0 & 0 & 0 & 0 & X_{11} & 0 & X_{12} & 0\\
 0 & 0 & 0 & 0 & 0 & 1 & 0 & 0\\
 0 & 0 & 0 & 0 & X_{21} & 0 & X_{22} & 0\\
 0 & 0 & 0 & 0 & 0 & 0 & 0 & 1\end{pmatrix}
\begin{pmatrix}
 1 & 0 & 0 & 0 & 0 & 0 & 0 & 0\\
 0 & 1 & 0 & 0 & 0 & 0 & 0 & 0\\
 0 & 0 & 0 & 0 & 0 & 0 & 0 & 0\\
 0 & 0 & 0 & 1 & 0 & 0 & 0 & 0\\
 0 & 0 & 0 & 0 & 1 & 0 & 0 & 0\\
 0 & 0 & 0 & 0 & 0 & 0 & 0 & 0\\
 0 & 0 & 0 & 0 & 0 & 0 & 1 & 0\\
 0 & 0 & 0 & 0 & 0 & 0 & 0 & 1\end{pmatrix} \; , 
\end{align}
We can see that this choice fulfills Eq.(\ref{eq:apply_LJ}), where 
$\hat{\mc{K}}_1$ takes care of column 3 and 6 while $\hat{\mc{K}}_2$ takes care 
of all the other columns. Of course,
\begin{equation}
X = \begin{pmatrix} X_{11} & X_{12}\\ X_{21} & X_{22}\end{pmatrix}
\end{equation}
must be a unitary matrix where all matrix elements have absolute value squared
equal to one half. For that we use the matrices defined in Eq.~(\ref{M:HADSYH}).
\end{widetext}

\section{Numerical simulation: Unraveling}
In this section we describe a simplified version of the unraveling method
presented in~\cite{BrePet02}. This method is easy to implement and with proper 
optimization reduces the computational costs of processing and memory.
\label{unrav}
The method consists of the following:
\begin{itemize}
    \item[I.] An initial state $\ket{\Psi_{\alpha}(0)}$ is taken from the set of 
        basis states with zero magnetization, i.e., half of spins up and half down.
    \item[II.] A spin is chosen randomly from the chain to which the local 
        operation (LO) is applied. The LO consists of an element of the set of 
        Kraus operators $\{\hat{\mc{K}}_r\}$. Then, with a probability 
        $q_r = \braket{\Psi_{\alpha}(0)}{\hat{\mc{K}}_r^\dagger\hat{\mc{K}}_r|\Psi_{\alpha}(0)}$ 
        the LO $\hat{\mc{K}}_r$ transform the initial state $\ket{\Psi_{\alpha}(0)}$ 
        as follows
        \[
            \ket{\Psi_{\alpha}(0)}\,\mapsto\,  \ket{\Psi_{\alpha}(1)} = \frac{\hat{\mc{K}}_r\ket{\Psi_{\alpha}(0)}}{\sqrt{q_r}}\,.
        \]
    \item[III.] With probability $p$ it is decided to apply a complete projective
        measurement. This measurement collapses the state $\ket{\Psi_{\alpha}(1)}$
        to one of the elements of the basis, which eliminates any trace of coherences
        created in the previous step.
\end{itemize}
At each discrete time $n=1, 2,\ldots$, steps II-III are repeated. The state 
of the system is saved at each Monte-Carlo (MC) time $t$, i.e., $t = N\cdot n$. The 
evolution ends at some time $t^*$ when the system reach the equilibrium subspace. 
Which is equivalent to $\ket{\Psi_{\alpha}(t^*)} = c_0(t^*)\ket{\bs{0}}+c_1(t^*)\ket{\bs{1}}$  
($\ket{\bs{0}}=\ket{00\cdots0}$ and $\ket{\bs{1}}=\ket{11\cdots1}$) with 
$|c_0(t^*)|^2+|c_1(t^*)|^2=1$. This protocol is repeated for each of the 
initial states $\{\ket{\Psi_{\alpha}(0)}\}_{\alpha=1,\ldots,N_{\textrm{sam}}}$, so that
the density matrix at each MC time is approximated as follows
\begin{align*}
    \varrho(t) &\approx \frac{1}{N_{\textrm{sam}}}\sum_\alpha \kb{\Psi_{\alpha}(t)}{\Psi_{\alpha}(t)}\,
\end{align*}

 \section{Estimation of the relaxation time}
 \label{rlxt}

For the classical evolution, we say that the system has been reached the 
equilibrium state when it is found in an ordered phase, in other words, all the 
spins are up or all down. Meanwhile, for the quantum evolution, it is necessary
to consider the equilibrium when the system is found in a superposition of the 
two possible states $\ket{\bs{0}}=\ket{00\cdots0}$ and 
$\ket{\bs{1}}=\ket{11\cdots1}$.
\begin{eqnarray}
     \ket{\Psi_{\alpha}(t)} = c_0(t)\ket{\bs{0}}+c_1(t)\ket{\bs{1}} \nonumber \\
    where \quad (|c_0(t)|^2+|c_1(t)|^2=1)\,.
\end{eqnarray}

The total number of elements in the ensemble can be written as $N_{{\rm sam}}= 
N_1+N_2+\cdots+N_{t_{\rm max}}$, where $N_t$ corresponds to the number of elements 
of the ensemble that reach the equilibrium subspace at MC time step $t$. So, we 
define the probability that the system reach the equilibrium subspace at time $t$ as
\[
    P_t = \frac{N_t}{N_{{\rm sam}}} 
\]

\begin{figure}[t]
    \includegraphics[width=0.46\textwidth]{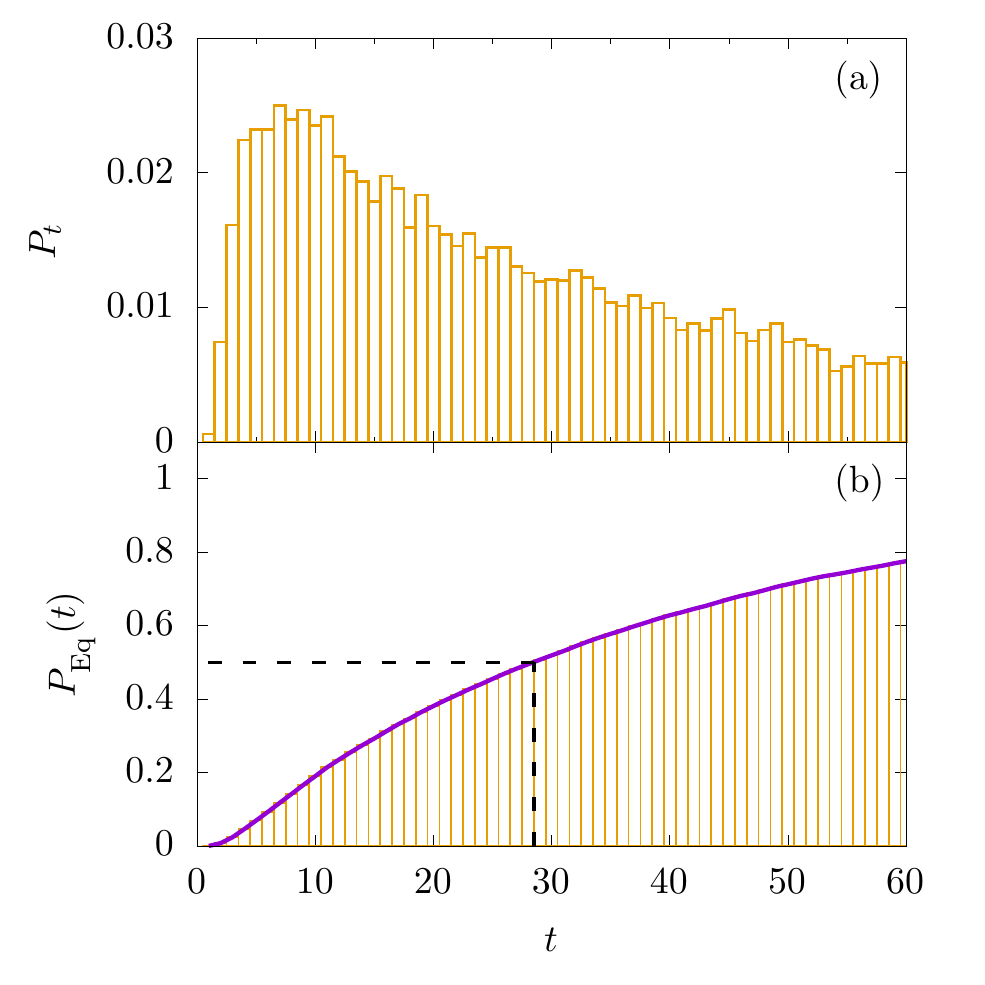}
    \caption{Panel a: Occurrence probability $P_t$ for a 16-spin chain and 
    "SYH" process. Panel b: Cumulative probability distribution calculated
    by the approximation of the integral (orange boxes). Alternatively, 
    calculated by means of $\tr[\hat{P}_{_{\rm Eq}}\varrho(t)]$ ( solid violet
    line). The dashed straight line indicates the value of $t_{1/2}$ obtained
    when $\hat{P}_{_{\rm Eq}}=1/2$.}
    \label{figappen}
\end{figure}
This probability of occurrence is plotted in figure \ref{figappen} (panel-a) for a chain
with 16-spin. 

Similarly we define the \textit{half-life} $t_{_{1/2}}$, the way it is defined in 
radioactive decay processes \cite{burton1960}, as the time 
required for half of the elements of the ensemble to reach the equilibrium 
subspace. To estimate the half-life as a function of the length of the spin 
chain (figure 2-b in the main text), we first calculate the cumulative 
probability distribution
\begin{equation}
    P_{{\rm Eq}}(t) = \int_0^t P_{t'}\, dt'
\end{equation}
and calculate the time for which $P_{{\rm Eq}}=1/2$, as shown in figure \ref{figappen} 
(panel-b, black dashed line). 

Alternatively, we can define an observable $\hat{\mathcal{O}}=\lambda_1
\hat{P}_{_{\rm Eq}} + \lambda_2 \hat{P}^c_{_{\rm Eq}}$ where 
$\hat{P}_{_{\rm Eq}}$ projects to the equilibrium subspace and 
$\hat{P}^c_{_{\rm Eq}}=\one-\hat{P}_{_{\rm Eq}}$. Then, the probability of 
reaching the equilibrium subspace $ P_{{\rm Eq}}(t)$ is given by 
\begin{equation}
    P_{{\rm Eq}}(t) = \tr[\hat{P}_{_{\rm Eq}}\,\varrho(t)]
\end{equation}
see figure \ref{figappen} (lower panel, solid violet line).

%
%
%


\begin{thebibliography}{99}
\bibitem{siml} Using a single XEON core with 512GB RAM, it takes 5-6 days to 
simulate a spin chain of 20 spins for 24000 realisations.
\end{thebibliography}
\end{document}